\documentclass{svproc}
\usepackage{url}
\usepackage{array}

\usepackage{graphicx}
\usepackage{multirow}
\begin{document}
\mainmatter

\title{Improved Processing of Ultrasound Tongue Videos by Combining ConvLSTM and 3D Convolutional Networks}
\titlerunning{Convolutional LSTM and 3D-CNN Networks}

\author{Amin Honarmandi Shandiz, L\'aszl\'o T\'oth}
\authorrunning{Amin Shandiz and L. Tóth}

\institute{University of Szeged, Institute of Informatics, Hungary \break
\email{\{shandiz,tothl\}@inf.u-szeged.hu}
}

\maketitle

\begin{abstract}
Silent Speech Interfaces aim to reconstruct the acoustic signal from a sequence of ultrasound tongue images that records the articulatory movement. The extraction of information about the tongue movement requires us to efficiently process the whole sequence of images, not just as a single image. Several approaches have been suggested to process such a sequential image data. The classic neural network structure combines two-dimensional convolutional (2D-CNN) layers that process the images separately with recurrent layers (eg. an LSTM) on top of them to fuse the information along time. More recently, it was shown that one may also apply a 3D-CNN network that can extract information along both the spatial and the temporal axes in parallel, achieving a similar accuracy while being less time consuming. A third option is to apply the less well-known ConvLSTM layer type, which combines the advantages of LSTM and CNN layers by replacing matrix multiplication with the convolution operation. In this paper, we experimentally compared various combinations of the above mentions layer types for a silent speech interface task, and we obtained the best result with a hybrid model that consists of a combination of 3D-CNN and ConvLSTM layers. This hybrid network is slightly faster, smaller and more accurate than our previous 3D-CNN model. 

\end{abstract}

\section{Introduction}
The area of Silent Speech Interfaces (SSI) deals with the problem of converting articulatory recordings to speech signals~\cite{schultz2017biosignal}. The studies in this field have considered various types of articulatory signals as input, such as Electroencephalography (EEG), Electromagnetic Articulography (EMA), Ultrasound Tongue Video Imaging (UTI) and so on. SSIs could provide a great amount of help for those disabled people who can not talk loud, but are able to silently articulate the speech. Converting the signals recorded during articulation to speech would allow these people to interact with others. SSI solutions could also be used in some other conditions where normal communication is not possible, for example in certain military situations or in very noisy industrial environments where people could barely understand each other. In this paper, we work with Ultrasound Tongue Video Imaging (UTI) Data~\cite{csapo2017dnn,toth20203d,jaumard2016articulatory} recorded from Hungarian speakers. 

In SSI systems, the classic approach of estimating the speech signal from the articulatory data consisted of two steps: the first one is to to convert the input to text using speech recognition, and the second is to synthesize speech based on the text. Nowadays, however, directly converting the articulatory signals to speech is more popular, as it is less time consuming and seems to be more suitable for real-time applications. This direct approach has been made viable by the Deep Neural Network (DNN) technology, which also revolutionized many other speech-related fields, for example speech recognition~\cite{hinton2012deep}, and speech synthesis~\cite{ling2015deep}. Here, we also follow the direct approach and apply DNNs~\cite{grosz2018f0,toth20203d,young2018recent}. 

The given articulatory-to-acoustic mapping task can be addressed by applying simple fully connected DNNs~\cite{grosz2018f0,jaumard2016articulatory}. However, as we are working with images, using convolutional neural networks (CNN)~\cite{krizhevsky2012imagenet} is more reasonable, and they have been successfully applied to the SSI task by several authors~\cite{saha2020ultra2speech,toth20203d}. A very important further aspect is that that our input consists of a sequence of images (an ultrasound video), so it is not effective to simply process these images separately. Thus, we can apply a Recurrent Neural Network (RNN) such as an LSTM~\cite{hochreiter1997long} to extract information from a sequence. And when the sequence consists of images, like in our case, an obvious solution is to combine an LSTM with a 2D CNN that extracts the information from individual video frames~\cite{juanpere2019ultrasound}. Alternatively, it would be possible to extend the 2D convolution to 3D by adding the time axis as an extra dimension~\cite{Tran}. This approach was followed in~\cite{toth20203d}, and it was found that the two approaches can achieve very similar performances. However, there is a third option: for the processing of image sequences, Shi et al proposed the ConvLSTM layer type~\cite{shi2015convolutional}, which combines the advantages of convolutional and recurrent processing in one layer. For some reason, however, the ConvLSTM construct is not widely known. To knowledge, it was applied to UTI data only in one case~\cite{zhao2019predicting}, but even in that paper the task was different from ours. 
The goal of this paper is to experiment with ConvLSTM models, and compare their performance with the previous 2D-CNN+LSTM and 3D-CNN approaches. We also try to combine the three types of layers, resulting in hybrid models, and our results show that the hybrid approach yields the best performance for our task.

The paper is organized as follows. In Section~\ref{convlstm}, we briefly introduce the concept of the Convolutional LSTM that we are going to use. In Section~\ref{data}, we explain the data acquisition and processing steps for our input and output data. In Section~\ref{setup}, we present our experimental setup, while in and following Section~\ref{result}, the experimental results are discussed and explained. Finally, in Section~\ref{conclusion} our main conclusions are given.

\section{Convolutional LSTM for SSI}
\label{convlstm}

 SSI systems synthesize speech from articulatory videos by learning the mapping between the input ultrasound image sequence and the output audio signal. SSI is a sequential task, as both the input and the output are sequences, with a strong correlation between consecutive elements of the sequence. As in our case the input data consist of ultrasound images, convolutional networks (CNN) seems to be a proper tool for processing the input, as they are known to perform well when working with images~\cite{krizhevsky2012imagenet}, and also in particular with SSI ultrasound tongue images~\cite{kimura2019sottovoce,juanpere2019ultrasound}. As our input is a sequence, the information content along the time axis of the data can be extracted by applying Recurrent Neural Networks (RNNs). 
 In particular, a variant of recurrent networks called the Long-Short Term Memory (LSTM) is known to be more effective in extracting long-term dependencies in the input sequence~\cite{hochreiter1997long}. 
 These networks have special gates in their internal implementations which improve their abilities to handle large-distance relations between time-related features. 
 
 \begin{figure}[!t]
\centering
\includegraphics[width=1.0\textwidth]{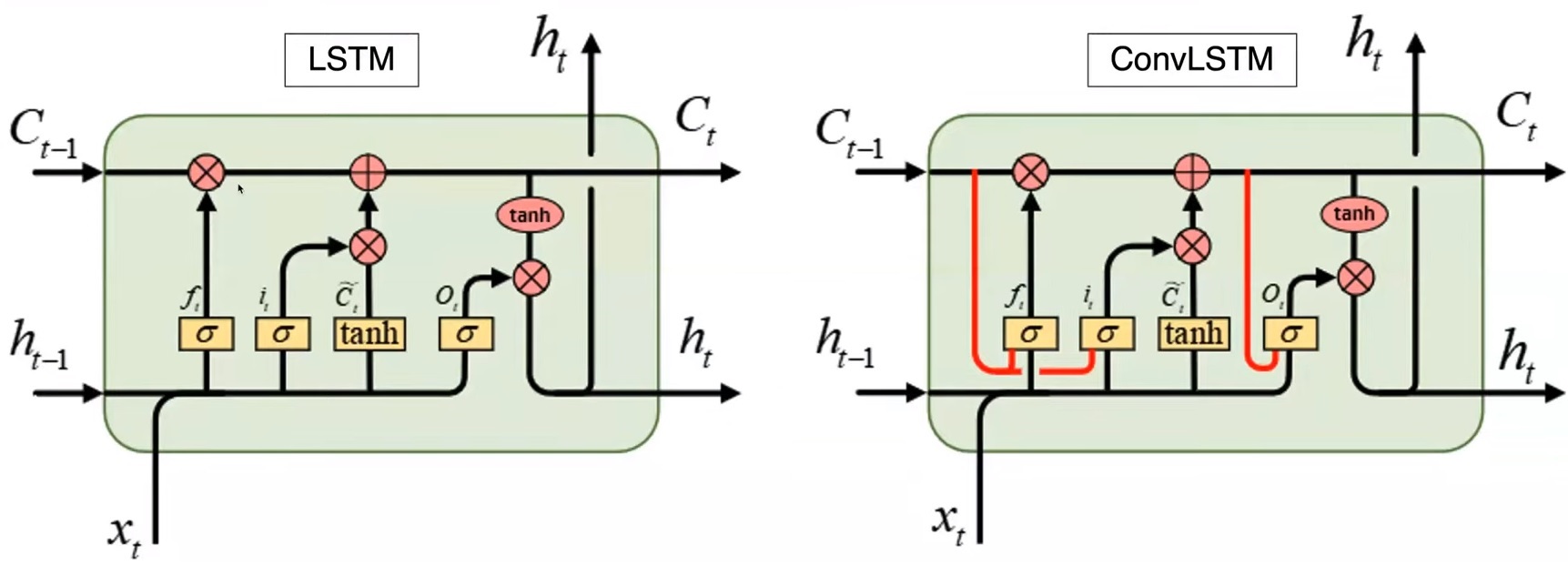}
\caption{\textit{Internal structure of a standard LSTM cell and its extended version (with extra peephole connections) used in Convolutional LSTMs~\cite{convlstm,RNNwithkeras}. }} \label{vs}
\end{figure}

\begin{figure}[!t]
\centering
\includegraphics[width=1.0\textwidth]{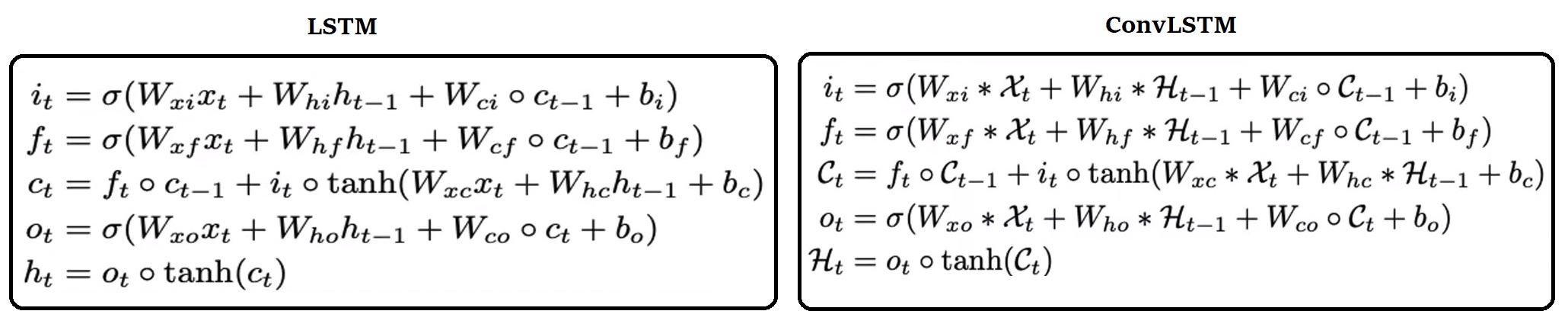}
\caption{\textit{The equations behind the operation of Long Short Term Memory (LSTM) versus Convolutional LSTM neurons~\cite{shi2015convolutional}. }} \label{vsequition}
\end{figure}

 The data flow in the standard implementation of an LSTM is shown in Fig.~\ref{vs}. Some implementations also contain extra connections, such a so-called "peephole" variant in shown on the right side of Fig.~\ref{vs}, In both cases, the input of the LSTM consist of a sequence of vectors. As in our case we want to process a sequence of images, a straightforward solution is to combine a CNN with an LSTM. In the trivial arrangement the images of the input are first processed by a (2-dimensional) CNN, and the sequence of CNN outputs are integrated over time by using an LSTM. This approach was shown to work fine in SSI implementations~\cite{juanpere2019ultrasound}. and we will shortly refer to this scheme as the CNN+LSTM approach. However, this type of processing requires the combination of two layers -- a 2D-CNN to process the data along the two spacial axes, and an LSTM to process it along the temporal axis. A more efficient solution called the 
 Convolutional LSTM, or shortly ConvLSTM has been proposed by Shi et al. As the name says, their solution performs the two processing steps in one. It can extract spatio-temporal features from the input data by applying a convolution operation in the inner steps LSTM instead of matrix multiplication~\cite{shi2015convolutional}. This is reflected in the equations of Fig.~\ref{vsequition}. where "$\ast$" represents the convolution operation, and "$\circ$" stands for gating. Note that, apart from the convolution, the equations are exactly the same as those of the (peephole) LSTM. The convolution allows the more efficient processing of image sequences, resulting in better performance with much fewer parameters. For example, Kwon et al. successfully applied a hierarchical ConvLSTM for speech recognition~
 \cite{kwon2020clstm}. Recently, Zhao et al. used ConvLSTMs for predicting subsequent ultrasound images in an SSI task~\cite{zhao2019predicting}. 
 
 Processing a sequence of images is also viable by extending the convolution operation to the time axis, resulting in a three-dimensional convolution (3D-CNN) model. The main advantage of this approach is that it is faster, as it applies only convolution operations. Convolution also allows the skipping of input images, which is not possible in an LSTM framework. In a previous paper, the 3D-CNN model gave results that were comparable or slightly better than those with the more conventional CNN+LSTM approach~\cite{toth20203d}. Here, we extend this earlier comparison to the ConvLSTM model, and we are also going to experiment with hybrid models that combine 3D-CNN and ConvLSTM layers.
 

\section{Data acquisition and preprocessing }\label{data}
The ultrasound data was collected from a Hungarian female subject (42 years old) while she was reading sentences aloud. Her tongue movement was recorded in a midsaggital orientation -- placing the ultrasonic imaging probe under the jaw -- using a "Micro" ultrasound system by Articulate Instruments Ltd. The transducer was fixed using a stabilizer headset. The 2-4 Mhz / 64 element 20 mm radius convex ultrasound transducer produced 82 images per second. The speech signals were recorded in parallel with an Audio-Technica ATR 3350 omnidirectional condenser microphone placed at a distance of 20 cm from the lips. The ultrasound and the audio signals were synchronized using the software tool provided with the equipment. Altogether 438 sentences (approximately half an hour) were recorded from the subject, which was divided into train, development and test sets in a 310-41-87 ratio. We should add that the same data set was used in several earlier studies \cite{csapo2017dnn,grosz2018f0,toth20203d}, and the data set is publicly available\footnote{The dataset is available upon request from csapot@tmit.bme.hu}. 
The ultrasound probe records 946 samples along each of its 64 scan lines. The recorded data can be converted to conventional ultrasound images using the software tools provided. However, due to its irregular shape, this image is harder to process by computers, while it contains no extra information compared to the original scan data. Hence, we worked with the original 964x64 data items, which were downsampled to 128x64 pixels. 
The intensity range of the data was min-max normalized to the [-1, 1] interval before feeding it to the network. 

The speech signal was recorded with a sampling rate of 11025 Hz, and then converted to a 80-bin mel-spectrogram using the SPTK toolkit (http://sp-tk.sourceforge.net). The goal of the machine learning step was to learn the mapping between the sequence of ultrasound images and the sequence of mel-spectrogram vectors. As the two sequences are perfectly synchronized, it was not necessary to apply a sequence-to-sequence learning strategy. We simply defined the goal of learning as an image-to-vector mapping task, using the mean squared error (MSE) as the loss function in the network training step. The 80 mel-frequency coefficients served as training targets, from which the speech signal was reconstructed using WaveGlow~\cite{waveglow}. To facilitate training, each of the 80 targets were standardized to zero mean and unit variance. The input of training consisted of a block of 25 consecutive. This allowed all DNN variants to involve the time axis in the information extraction process. The whole SSI framework followed our earlier study~\cite{toth20203d}.

\section{Experimental Setup}\label{setup}
We implemented our networks using Keras with a tensorflow back-end~\cite{chollet2015keras}. We applied three different network architectures that can process 3-dimensional blocks of data. In the tables, "3D-CNN" refers to the fully convolutional model proposed in~\cite{toth20203d}. This model does not have any LSTM component. "3D-CNN + BiLSTM" refers to a combination which applies a BiLSTM layer as the topmost hidden layer to integrate the temporal features extracted by the previous 3D-CNN layers. The final model referred to as "3D-CNN + ConvLSTM" replaces the tompost 3D-CNN and BiLSTM layers by a ConvLSTM layer. Notice that the ConvLSTM technique fuses the convolution and the LSTM operations into one layer, so here we can also spare one hidden layer by this substitution. In the following, we give a more detailed description of the three configurations.

\textbf{3D Convolutional Neural Network(3D-CNN):} This model was described in detail in~\cite{toth20203d}, and its network layers are shown in Table~\ref{tab1}. The networks processes the input sequence of 25 video frames in 5-frame blocks using 3D convolution. The overlap between these blocks is minimized by setting the stride parameter $s$ of the time axis to 5. These blocks are processed further by 3 additional Conv3D layers, with pooling layers after every second convolution layer. Finally, the output is flattened and integrated over the time axis by a dense layer as the topmost hidden layer. The output hidden layer is a linear layer with 80 neurons, corresponding to the 80 spectral parameters given as training targets. This special network structure was motivated by Tran et al., who found that for the best result the processing should focus on the two spacial axes first, performing the integration over the temporal axis only afterwards~\cite{Tran}. Toth at al. also obtained the best result with performing the 3D convolution in this decomposed, "(2D+1)" form~\cite{toth20203d}. Compared to that study, we achieved slightly better results with the same architecture by switching to the Adam optimizer instead of SGD, and by adjusting some meta-parameters, for example the dropout rate.

\textbf{3D CNN + BiLSTM:} As the output of the four layers of 3D convolution, the 3D-CNN network produces a sequence of 5 matrices, which are combined by a simple dense layer (cf. Table~\ref{tab1}). Our first modification was to replace this fully connected layer with a (bidirectional) LSTM, which required us to reshape the matrices into vectors. The LSTM is a more sophisticated solution to extract the information from a temporal sequence, so we hoped to get slightly better results from this approach. As Table~\ref{tab1} shows, we set the $return\_sequences$ parameter of the LSTM to False, so the output is a simple vector, which serves as the input of the subsequent dense linear output layer.

\textbf{3D CNN + ConvLSTM:} Our main goal in this paper was to examine the efficiency of the ConvLSTM layer for this task. In the first experiment we applied it only at the topmost hidden layer of our 3D-CNN model (see Table~\ref{tab1}). As the ConvLSTM layer implements the operation of a convolutional and an LSTM layer in one, we replaced the uppermost Conv3D and LSTM layers by a ConvLSTM layer, reducing the number of neural hidden layers from 5 to 4. Also, as the ConvLSTM layer works with matrices and also outputs matrices, the reshaping was required after the layer and not before it.

\begin{table}[ht]
   \caption{The layers of the 3D-CNN, the 3D-CNN + BiLSTM and the 3D-CNN + ConvLSTM networks in Keras implementation, along with their most important parameters. The differences are highlighted in bold.}
   \centering
\renewcommand{\arraystretch}{1.1} 
\begin{tabular}{|l|l|}
\hline
\hfil3D-CNN & \hfil3D-CNN + BiLSTM  \\
\hline
Conv3D(30, (5,13,13), strides=(s, 2,2)) & Conv3D(30,(5,13,13), strides=(s,2,2))\\
Dropout(0.3) & Dropout(0.3)\\
\hline
Conv3D(60, (1,13,13), strides=(1,2,2)) & Conv3D(60,(1,13,13),strides=(1,2,2))\\
Dropout(0.3) & Dropout(0.3)\\
MaxPooling3D(pool\_size=(1,2,2)) & MaxPooling3D(pool\_size=(1,2,2))\\
\hline
Conv3D(90, (1,13,13), strides=(1,2,1)) & Conv3D(90,(1,13,13),strides=(1,2,1))\\
Dropout(0.3) & Dropout(0.3)\\
\hline
Conv3D(85, (1,13,13), strides=(1,2,2))~~~ &Conv3D(85, (1,13,13), strides=(1,2,2))\\
Dropout(0.3) & Dropout(0.3)\\
MaxPooling3D(pool\_size=(1,2,2)) & MaxPooling3D(pool\_size=(1,2,2))\\
\hline
\textbf{Flatten()} & \textbf{Reshape((5, 340))}\\
\textbf{Dense(500)} & \textbf{Bidirectional(LSTM(320,}\\
Dropout(0.3) & \textbf{~~~~~~~~~~~~~~~~~~~~~ret\_seq=False))}\\
\hline
Dense(80, activation='linear') & Dense(80, activation='linear')\\
\hline
\end{tabular}
\vspace{0.6cm}\newline
\begin{tabular}{|l|}
\hline
\hfil3D-CNN + ConvLSTM  \\
\hline
Conv3D(30,(5,13,13),strides=(s,2,2)) \\
Dropout(0.35)\\ 
\hline
Conv3D(60,(1,13,13),strides=(1,2,2))\\ 
Dropout(0.35) \\
MaxPooling3D(poolsize =(1,2,1))\\
\hline
Conv3D(90,(1,13,13),strides=(1,2,2)) \\
Dropout(0.35)\\
\hline
\textbf{ConvLSTM2D(64, (3,3), Strides=(2,2), ret\_seq=False)} \\
\textbf{Flatten()}\\
\hline
Dense(80,activation=’linear’)\\
\hline
\end{tabular}
\label{tab1}
\end{table}

\section{Results and Discussion}\label{result}
In the first experiment we compared the performance of the baseline 3D-CNN model we the two hybrid solutions proposed in the previous chapter and in Table~\ref{tab1}. In Table~\ref{tab2} we report two simple objective metrics of the quality of training, the mean squared error (MSE) and the $R^2$ score, which is popular in regression tasks implemented with neural networks (for $R^2$ a higher value means better performance). As can be seen, replacing the dense layer by the LSTM layer already brings a slight but consistent improvement in the results, both on the development and on the test set. Fusing the uppermost Conv3D and the LSTM layer into a ConvLSTM layer resulted in further error reduction of about the same rate, even though the network depth is decreased. This clearly proves the efficiency of the ConvLSTM layer. However, we also observed a drawback, namely that the ConvSLTM layer has much more trainable parameters than the Conv3D layer. Hence, we had to reduce the filter size in the ConvLSTM layer, in order to keep the number of parameters in the original range. Theoretically, similar to the LSTM layer, the ConvLSTM layer can also be made bidirectional. However, we ran into the same problem 
that it tremendously increased the number of parameters while yielding only a marginal improvement. Thus, we stuck with using the unidirectional variant. Finally, to fuse the Conv3D and the LSTM layers, we had to remove the second MaxPooling layer. We also tried to insert it back after the ConvLSTM layer, but the results did not change considerably.

\begin{table}[ht]
\caption{The MSE and mean $R^2$ scores obtained with the various network configurations for the development and test sets, respectively. The best results are highlighted in bold.}\label{tab2}
\centering
\renewcommand{\arraystretch}{1.3} 
\begin{tabular}{|l|c|c|l|c|}
\hline
\multicolumn{1}{|c|}{\multirow{2}{*}{Network Type}} & \multicolumn{2}{c|}{Dev} & \multicolumn{2}{c|}{Test} \\ \cline{2-5} 
\multicolumn{1}{|c|}{}                              & ~~~MSE~~~        & ~Mean $R^2$~     & ~~~MSE~~~        & ~Mean $R^2$~      \\ \hline
~3D-CNN                                         & 0.292      & 0.714       & 0.293      & 0.710         \\ \hline
~3D-CNN + BiLSTM~                                     & 0.285      & 0.721       & 0.282      & 0.721        \\ \hline
~\textbf{3D-CNN + ConvLSTM}~                                   & \textbf{0.276}      & \textbf{0.727}        & \textbf{0.276}      & \textbf{0.73}        \\ \hline
\end{tabular}

\end{table}

\begin{table}[ht]
    \caption{The MSE for different combinations of Conv3D and ConvLSTM layers in the four hidden layers of the network. The best results are highlighted in bold. }
    \centering
    \renewcommand{\arraystretch}{1.3} 
    \begin{tabular}{|l|l|l|l|l|l|l|}
    \hline
    Layer 1 & Layer 2 & Layer 3 & Layer 4 & ~~Dev~~ & ~~Test~~ \\ 
    \hline\hline
          ~ConvLSTM~ &~ConvLSTM~ &~ConvLSTM~ &~ConvLSTM~ & 0.31 & 0.31\\
         \hline
         ~ConvLSTM &~ConvLSTM &~ConvLSTM & ~~~~~~--- & 0.29 & 0.3\\
         \hline
          \hline
         ~Conv3D & ~ConvLSTM &~ConvLSTM & ~ConvLSTM & 0.31 & 0.31 \\
         \hline
          ~Conv3D~ &~Conv3D& ~ConvLSTM & ~ConvLSTM & 0.36 & 0.35\\
          \hline
          ~\textbf{Conv3D} &~\textbf{Conv3D} &~\textbf{Conv3D} & ~\textbf{ConvLSTM}~ & \textbf{0.27} &\textbf{ 0.27} \\
                   \hline
\hline
          ~Conv3D & ~ConvLSTM &~Conv3D & ~ConvLSTM & 0.3 & 0.3 \\
         \hline
          ~ConvLSTM &~Conv3D & ~Conv3D & ~ConvLSTM & 0.34 & 0.34\\
          \hline

    \end{tabular}
    
    \label{tab3}
\end{table}

Obviously, many other possible configurations exist that combine Conv3D and ConvLSTM layers.
In the second experiment we tried out further combinations of these two layers. We experimented with 4 hidden layer constructs, and we fixed the uppermost layer to be a ConvLSTM, as it convincingly proved to be the better setup in the previous experiment. Table~\ref{tab3} summarizes the architectures we experimented with. As regards ConvLSTM layers, the $return\_sequences$ parameter was set to True for intermediate layers, and set to False only when the ConvLSTM layer was the topmost hidden layer. The meta-parameters were always chosen so that the global count of the free parameters stayed similar to that of the baseline model.

Seeing the good performance of the ConvLSTM layer in the previous experiment, we first tried to build a fully ConvLSTM model. However, as the first row of Table~\ref{tab3} shows, we obtained no improvement. As the ConvLSTM layer proved to be more efficient than the Conv3D layer earlier, next we tried to create a network of just 3 ConvLSTM layers instead of 4. As shown in the second row of the table, the results became slightly better, but still worse than the baseline. This result reinforces our previous observation that ConvLSTM networks do not require the same depth as a convolutional network.

Conv3D layers have the advantage that they can easily downsample the time axis using a stride parameter larger than 1. On the contrary, ConvLSTM units cannot easily skip elements of their input sequence, due to their recurrent nature, which results in a large parameter count and a slow training. Hence, it seemed to be more efficient to put a Conv3D layer into the first hidden layer. We tried to place 1-2-3 Conv3D layers in the lower layers, and 3-2-1 ConvLSTM  layers in the remaining layers. The middle block of Table~\ref{tab3} shows that the optimal solution is to have just one ConvLSTM layer, as in our original experiment.
Lastly, we tried two further configurations with alternating Conv3D and ConvLSTM layers, motivated by papers like~\cite{kwon2020clstm,zhao2019predicting}, but we did not receive any better results. 

\section{Conclusion}\label{conclusion}
Here, we were seeking the optimal neural network architecture for the articulatory-to-acoustic mapping task of SSI systems. The task involves the processing of 3D data blocks -- sequences of images -- for which one can apply 3D-CNN models, such as in~\cite{toth20203d}. Alternatively, one may apply a ConvLSTM model proposed by~\cite{shi2015convolutional}. Besides comparing the purely convolutional and ConvLSTM models, we also experimented with hybrid architectures where the two layers types are mixed. The 3D-CNN + ConvLSTM hybrid model obtained the best results, better than the baseline 3D-CNN model, and it also outperformed other models with a different order of layers, as applied in~\cite{kwon2020clstm} for emotion recognition, and in~\cite{zhao2019predicting} for the prediction of the subsequent ultrasound image. Applying the ConvLSTM layer in the uppermost hidden layer even made the model smaller (with one hidden layer) and slightly faster to train. The optimal model arrangement consists of three Conv3D layers and a ConvLSTM on top of them, which illustrates that it is worth combining the ConvLSTM layer with other layer types such as the Conv3D to extract spatio-temporal features from videos -- in our case, to better capture the tongue movement. The winning architecture also shows that the Conv3D blocks are more efficient in extracting local spectro-temporal information, while ConvLSTM is more efficient in fusing these pieces of information along the time axis. Interestingly, this coincides with the observation of Tran et al. about the optimal order of feature extraction for 3D video blocks~\cite{Tran}. In the future we plan to extend our research to transformer models that apply two separate networks for encoding and decoding, which would allow us to experiment with different network types for the decoder and encoder components, similar to the UNET~\cite{behboodi2019ultrasound} architecture.

\section{Acknowledgments}

Project no. TKP2021-NVA-09 has been implemented with the support provided by the Ministry of Innovation and Technology of Hungary from  the National Research, Development and Innovation Fund, financed under the TKP2021-NVA funding scheme, and also within the framework of the Artificial Intelligence National Laboratory Programme. The RTX A5000 GPU used in the experiments was donated by NVIDIA.

\bibliographystyle{splncs}
\bibliography{convlstm}
 
\end{document}